\documentclass[12pt,preprint]{aastex}

\usepackage{emulateapj5}

\slugcomment{Accepted for the Astrophysical Journal}

\shorttitle{Warm Spitzer Photometry of CoRoT Planets}
\shortauthors{Deming et al.}

\begin{document}


\title{{\it Warm Spitzer} Photometry of the Transiting Exoplanets CoRoT-1 
and CoRoT-2 at Secondary Eclipse}


\author{Drake~Deming\altaffilmark{1}, 
Heather Knutson\altaffilmark{2,3}, Eric Agol\altaffilmark{4}, Jean-Michel Desert\altaffilmark{5}, 
  Adam Burrows\altaffilmark{6},
Jonathan~J.~Fortney\altaffilmark{7}, David Charbonneau\altaffilmark{5}, 
Nicolas~B.~Cowan\altaffilmark{4,8}, 
Gregory Laughlin\altaffilmark{7},
Jonathan Langton\altaffilmark{9}, Adam~P.~Showman\altaffilmark{10}, 
and Nikole~K.~Lewis\altaffilmark{10}}


\altaffiltext{1}{Planetary Systems Laboratory, NASA's Goddard Space Flight Center, 
Greenbelt MD 20771}
\altaffiltext{2}{Department of Astronomy, University of California at Berkeley, Berkeley CA 94720}
\altaffiltext{3}{Miller Research Fellow}
\altaffiltext{4} {Department of Atronomy, University of Washington, Box 351580, Seattle, WA 98195}
\altaffiltext{5}{Harvard-Smithsonian Center for Astrophysics, Cambridge, MA 02138}
\altaffiltext{6}{Department of Astrophysical Sciences, Princeton University, Princeton, NJ 05844}
\altaffiltext{7}{Department of Astronomy and Astrophysics, University of California at Santa Cruz, 
 \\ Santa Cruz, CA 95064}
\altaffiltext{8}{Currently: CIERA Fellow, Department of Physics \& Astronomy, Northwestern University,
    2131 Tech~Drive, Evanston, IL 60208} 
\altaffiltext{9} {Department of Physics, Principia College, Elsah, IL 62028}
\altaffiltext{10} {Lunar and Planetary Laboratory, University of Arizona, Tucson, AZ 85721}

\begin{abstract}
We measure secondary eclipses of the hot giant exoplanets CoRoT-1 at
3.6 and 4.5\,$\mu$m, and CoRoT-2 at 3.6\,$\mu$m, both using Warm
Spitzer. We find that the Warm Spitzer mission is working very well
for exoplanet science.  For consistency of our analysis we also
re-analyze archival cryogenic Spitzer data for secondary eclipses of
CoRoT-2 at 4.5 and 8\,$\mu$m.  We compare the total data for both
planets, including optical eclipse measurements by the CoRoT mission,
and ground-based eclipse measurements at 2\,$\mu$m, to existing
models. Both planets exhibit stronger eclipses at 4.5 than at
3.6\,$\mu$m, which is often indicative of an atmospheric temperature
inversion.  The spectrum of CoRoT-1 is best reproduced by a 2460K blackbody,
due either to a high altitude layer that strongly absorbs stellar
irradiance, or an isothermal region in the planetary atmosphere. The
spectrum of CoRoT-2 is unusual because the 8\,$\mu$m contrast is
anomalously low.  Non-inverted atmospheres could potentially produce
the CoRoT-2 spectrum if the planet exhibits line emission from CO at
4.5\,$\mu$m, caused by tidal-induced mass loss. However, the viability
of that hypothesis is questionable because the emitting region cannot
be more than about 30\% larger than the planet's transit radius, based
on the ingress and egress times at eclipse.  An alternative
possibility to account for the spectrum of CoRoT-2 is an additional
opacity source that acts strongly at wavelengths less than 5\,$\mu$m,
heating the upper atmosphere while allowing the deeper atmosphere seen
at 8\,$\mu$m to remain cooler.  We obtain a similar result as
\citet{gillon10} for the phase of the secondary eclipse of CoRoT-2,
implying an eccentric orbit with $e\,cos(\omega)=-0.0030\pm0.0004$.
\end{abstract}



\section{Introduction}

An especially interesting class of giant extrasolar planets, the `very
hot Jupiters' (hereafter, VHJs), orbit extremely close to solar-type
stars, within 0.03 AU in several cases.  The temperature structure in
the atmosphere of such a planet is likely to be significantly
perturbed by the strong stellar irradiation.  Absorption of stellar
radiation is one possible energy source that may drive atmospheric
temperature inversions. Temperature inversions with height appear to
be common in hot Jupiter atmospheres; they occur over a wide range of
stellar irradiation level \citep{knutson08, machalek, christiansen,
todorov}, but are not well understood.  The emergent spectra of VHJs
are an important key to this problem.  The emergent spectrum of a
transiting planet can often be measured by observing the decrease in
total light as the planet passes behind the star during secondary
eclipse \citep{charb05, deming05}. Eclipses of the VHJs offer the
opportunity to determine their emergent spectra at wavelengths as
short as visible light \citep{alonso09a, snellen09}. Fortunately, VHJs
have high transit probabilities, and are represented by transiting
planets such as WASP-12 \citep{hebb09}, WASP-19 \citep{hebb10},
CoRoT-1 \citep{barge}, and CoRoT-2 \citep{alonso08}.

The CoRoT planets are particularly important in the study of VHJ
temperature structure. Their emergent flux has been measured at
secondary eclipse using infrared (IR) wavelengths, and also in the
visible by the CoRoT mission. The currently available secondary
eclipse measurements for the CoRoT planets are summarized in Table~1,
including the results from this paper.  While eclipses of CoRoT-2 have
been measured at 4.5\,$\mu$m and 8.0\,$\mu$m using the Spitzer
Space Telescope \citep{gillon10}, no Spitzer measurements have
been reported for CoRoT-1.  In this paper, we report measurements of
CoRoT-1 using Warm Spitzer \citep{deming07} at $3.6$- and
4.5\,$\mu$m, and we complete Spitzer's measurement of CoRoT-2 by
adding the 3.6\,$\mu$m observation.  These additional data allow us to
address the existence and nature of the inversion phenomenon in these
planets.  Moreover, because we measure near the peak of the VHJ's
spectral energy distribution, we can speak to whether the visible
wavelength eclipse measurements are sensing primarily thermal
radiation, as opposed to reflected light. 

Our results, together with those of \citet{hebrard}, are among the
first to be reported for transiting exoplanets using Warm Spitzer.
The Warm phase of Spitzer refers to operation of the observatory after
the loss of cryogen, with only the 3.6- and 4.5\,$\mu$m channels of
the IRAC instrument remaining operational.  The InSb detectors used at
these wavelengths are now functioning at a temperature of
approximately 29 Kelvins, cooled by passive radiation. This very
different operating temperature regime may have significant
implications for the observatory performance as regards high precision
photometry. Therefore, we comment on the performance of the
observatory, within the limits allowed by the fact that we have
observed relatively faint stars.

In Sec.~2 we describe the observations, aperture photometry, and
linear regression procedures to derive eclipse depths and central
phases.  Sec.~3 discusses the implications of our results for the
orbital and atmospheric properties of these giant CoRoT planets, and
in the Appendix we discuss some details concerning the performance of the Warm
Spitzer observatory for this type of exoplanet science.

\section{Observations and Photometry}

\subsection{CoRoT-1}

We observed CoRoT-1 at 4.5\,$\mu$m on 23 November 2009, starting at
11:06 UT (orbital phase 0.380), for a duration of 465.7 minutes,
yielding 888 30-second exposures. Among transiting systems, CoRoT-1 is
relatively faint, having V=13.6 and K=12.1, and a short orbital period
of P=1.509 days \citep{barge}. We observed this system at 3.6\,$\mu$m
on 26 November 2009, starting at 11:30 UT (orbital phase 0.379) for
the same duration, and the same exposure time per frame.  The CoRoT-1
observations at both wavelengths used full frame ($256\times
256$-pixel) mode.  Following the eclipse observations, we acquired 9
minutes of additional data (17 exposures) by offsetting the telescope
to view blank sky using the same detector pixels as for CoRoT-1.

The detectors in the Warm mission are significantly affected by an
artifact called column pull-down\footnote{see
http://ssc.spitzer.caltech.edu/irac/warmfeatures/}, wherein the
presence of a bright star reduces the signal level for an entire
detector column. This, as well as other artifacts, are significantly
mitigated in the cBCD files produced by Spitzer's pipeline processing.
However, neither CoRoT-1 nor CoRoT-2 lie on columns affected by
pull-down, and in any case we would want to remove any such artifacts
as an integral part of our photometry, so that we could fully judge
their impact. We therefore extracted photometry using the Basic
Calibrated Data (BCD) files produced by version S18.12.0 of the
Spitzer pipeline, not the cBCD files. We calculated orbital phase
using the UTC-based HJD values for the start of each observation from
the FITS headers of the BCD files, and we correct the values to the
time of mid-exposure.

As a first step, we stack the blank sky images and median-filter each
pixel in time to construct an average blank sky frame. We subtract
this sky frame from each CoRoT-1 image immediately after reading each
BCD file. In principle, this subtraction of a sky-nod will remove the
background radiation, but we nevertheless fit and remove residual
background anyway, as described below. Although the true sky
background should be constant to an excellent approximation, we find
that the background does vary significantly from frame to frame.  This
is one significant difference from the cryogenic mission, as we
discuss in the Appendix.

We locate and correct energetic particle events by comparing the
time history of each pixel to a 5-point median filter of that pixel
intensity {\it vs.} time, and we replace $>4\sigma$ outliers with the
median value. The fraction of pixels we correct varies between 0.45\%
and 0.55\%, depending on which planet and wavelength are analyzed. We
perform aperture photometry on the images, after first applying
corrections for variations in pixel solid angle, and for slightly
different flat-field response for point sources {\it vs.}  extended
sources\footnote{see Secs. 5.3, and 5.6.2 of the IRAC Data Handbook,
V3.0}. Prior to subtracting the residual background and performing
aperture photometry, we convert the pixel intensities to electrons,
using the calibration information given in the FITS headers. This
facilitates the evaluation of the photometric errors.

Our photometry code locates the centroid of the stellar point spread
function (PSF) by fitting a symmetric 2-D Gaussian to the PSF-core
\citep{agol}.  We calculate the flux within a centered circular
aperture, of variable radius, using radii of 2.0 to 4.5 pixels, in
0.5-pixel steps.  To determine the residual background intensity, we
fit a Gaussian to a histogram of pixel intensities for each frame.
The center histogram bin, defined to fractional precision by the
Gaussian fit, is adopted as the residual background intensity.
Subtracting the resultant background from the raw aperture photometry,
yields 6 photometric time series for the star corresponding to
aperture radii from 2.0 to 4.5 pixels.  We tabulate the magnitude of
the point-to-point scatter in our photometry, and errors in our final
results, as a function of aperture radius.  We find that both the
scatter and final parameter errors depend only weakly on aperture
radius, with best values near 2.5 to 3.0 pixels.  We adopt a radius
of 3.0 pixels for all of our photometry.

The aperture photometry for CoRoT-1 at 3.6\,$\mu$m, uncorrected for
instrument systematic effects, is shown in the top panel of Figure~1.
The corresponding time series at 4.5\,$\mu$m is shown in the top panel
of Figure~2.

\subsection{CoRoT-2}

CoRoT-2 (V=12.6, K=10.3) observations at 3.6\,$\mu$m began on 24
November 2009 at 18:22 UT (orbital phase 0.4), for a duration of 467.6
minutes. CoRoT-2 being brighter than CoRoT-1, these observations used
subarray mode. We collected 215 data cubes, each comprising $64$
2-second exposures of $32\times32$ pixels, followed by 3 data cubes of
blank sky.

We perform photometry on the CoRoT-2 data cubes in a similar manner to
the full-frame data for CoRoT-1. We inspect the aperture photometry
for the 64 frames within each data cube, and zero-weight outliers
exceeding the average by more than $4\sigma$.  The first frame in each
data cube is consistently found to be an outlier, and is always
ignored.  We analyze the remaining 63-frame data cubes so as to
produce two distinct versions of the photometry, and we perform the
entire eclipse-fitting and error analysis for each version.  In the
first (default) version, we average the background-subtracted aperture
photometry for all 63 frames in each data cube, to produce a single
photometric point.  For the second version, we use each of the 63
frames as a separate photometric point. Using these individual frames
potentially exploits the short-term pointing jitter to better define
the intra-pixel effect.  However, in practice the frame-to-frame
fluctuations within a data cube are dominated by photon noise for
these relatively faint stars. The eclipse results and errors from
these two versions of the photometry are close to being identical
(difference much less than $1\sigma$).  Note that the default method
is essentially just a binning of the data.  We prefer the default
version because the eclipse plot (Figure~3) is visually clearer. 

We also explored a third version of the photometry, wherein we average
the actual data frames in each data cube, omitting the first frame and
using a median filter to reject outlying pixels.  We then perform
aperture photometry on the averaged frame.  This method gives
essentially the same result as our default method: the eclipse
amplitude (see below) differed by 0.4$\sigma$ and the phase differed
by 1.14$\sigma$.

\subsection{Eclipse Amplitudes}

CoRoT-1 and -2 have well defined transit parameters (planetary and
stellar radii, orbit inclination, etc.).  We adopt these parameters
from \citet{barge} and \citet{alonso08}, and we calculate eclipse
curves numerically, following \cite{todorov}.  We maintain the
known durations of ingress and egress, but we vary the central phase and
eclipse depth when fitting to the photometry.

Both the 3.6- and 4.5\,$\mu$m channels show the well known intra-pixel
sensitivity variation \citep{morales}. We fit for the eclipse depth
and the coefficients of the intra-pixel correction using linear
regression. The details of the fitting procedure vary with wavelength,
but at all wavelengths we search for the best central phase by
repeating the linear regressions at many phase values in a dense grid
(spacing 0.0002 in phase), and we adopt the central phase yielding the
best $\chi^2$.  We always perform this grid-search in phase when
fitting for eclipse amplitude, for both planets at all wavelengths and
also for our Monte-Carlo trials to define errors (see below).

We apply the linear regressions using an iterative procedure. We first
decorrelate the photometry to remove the intra-pixel effect, while
ignoring the eclipse, and then we fit for the eclipse depth using a
second regression on the decorrelated data.  After removing the fitted
eclipse depth from the original photometry, we then re-fit and
decorrelate the intra-pixel variation, then re-fit the eclipse.  This
procedure converges in two cycles.  In principle, iteration is
unnecessary because the regressions are linear and an identical result
can be achieved by solving simultaneously for both the intra-pixel
coefficients and the eclipse depth.  (We verified this by actually
doing the simultaneous fit for a simple case.)  Nevertheless, we use
the iterative procedure because in actual practice it is more flexible
and it affords the opportunity to use variants of the fit that would
be awkward to implement in a simultaneous solution.  This should
become apparent from the description below.

At 3.6\,$\mu$m the intra-pixel signature in the photometry ($\sim 2\%$
peak-to-peak) is larger than the eclipse (see top panel of Figure 1).
Our first step is to solve for a provisional intra-pixel
decorrelation.  The provisional decorrelation function is assumed to
be linear in both $\delta X$ and $\delta Y$, which are defined as the
change in X- and Y-pixel position of the image centroid after removing
a trend in $X$ and $Y$ with time.  The approximately 1-hour
quasi-periodic jitter in position has peak-to-peak amplitude in
$\delta X$ and $\delta Y$ of about 0.03 and 0.08-pixels, respectively.
The trends (slow drifts) are smaller, about 0.005-pixels in X over the
entire dataset, and 0.06-pixels in Y.  The provisional intra-pixel
decorrelation function is linear in both $\delta X$ and $\delta Y$,
and includes a term linear in time that accounts for both the slow
drift in position as well as possible change in detector
sensitivity. We solve for the coefficients using linear regression
(matrix inversion), and correct the original photometry using this
decorrelation function.

Following the provisional intra-pixel decorrelation, we solve for the
eclipse depth, again using linear regression.  This regression
formally allows a linear baseline in time, but that term is
effectively accounted for by the intra-pixel decorrelation of the
previous step.  We remove the fitted eclipse from the original
photometry, and begin the second cycle of the iteration. This
implements a more sophisticated version of the intra-pixel
decorrelation, expressing the decorrelation function as linear in both
time and the radial distance of the image from pixel center (called
pixel phase).  Because there is slow drift of the image toward pixel
center by about $0.06$ pixels over the duration of the observations,
intrinsic spatial variation in the intra-pixel sensitivity (i.e., a
change of spatial slope) may be manifest as a change in the
decorrelation coefficient of pixel phase.  In this particular case
(CoRoT-1 at 3.6\,$\mu$m), visual inspection of the data indeed
suggested a change in the slope of the intrapixel effect. To allow for
this change in slope, we divide the decorrelation into two halves, the
first half before mid-eclipse and the second half after mid-eclipse.
In effect, this is a minimalist implementation of using a quadratic
term in the intrapixel fit. Although it is unconventional, we judge it
to be the best approach to this particular case.  The coefficients of
both halves are found via linear regression on the eclipse-removed
data.  The separate decorrelation functions for the first and second
halves of the data can be discerned on the top panel of Figure~1.
Note that they are almost continuous at the break near phase 0.5.
None of the conclusions of this paper would be different if we
restricted the decorrelation to more conventional methodology, but the
quality of the 3.6\,$\mu$m eclipse fit for CoRoT-1 would be degraded.

After this decorrelation, we again remove the intra-pixel effect from
the original photometry, and re-solve for the final eclipse depth and
a possible linear baseline via regression.  The eclipse fit uses all
of the data, not breaking it into halves. Decorrelated CoRoT-1 data
and the best-fit eclipse are shown in the middle panel of Figure~1, and
are binned (to 100 bins) in the bottom panel of Figure~1.

We use a nearly identical procedure to fit the 3.6\,$\mu$m eclipse of
CoRoT-2, shown in Figure~3, except that we do not break the
decorrelation at mid-eclipse.  The first $\sim 30$ minutes of these data
(not illustrated in Figure~3) exhibit a transient decrease in flux,
similar to the ramp effect seen at longer wavelength, but decreasing
instead of increasing, and not correlated with the image position on
the detector. Transient effects at this wavelength are not well
understood, so we simply omit the 19 data cubes prior to orbital
phase 0.41.

Another difference for CoRoT-2 is that a correction is needed for
diffracted light from an M-dwarf lying 4 arc-sec distant
\citep{gillon10}. Since we also re-analyze archival data at 4.5 and
8\,$\mu$m for CoRoT-2 (see below), we need to estimate the diffracted
light contributed by the M-dwarf in the CoRoT-2 aperture at 3.6, 4.5,
and 8\,$\mu$m.  We calculated the flux ratio (M-dwarf to CoRoT-2) in
the IRAC bands, using the the flux estimation tool (STAR-PET) on the
Spitzer website, and the 2MASS K-magnitudes and J-K colors of the two
stars.  Knowing their relative brightness, we also need to know the
fraction of the M-dwarf flux that is diffracted into the photometry
aperture for CoRoT-2.  We estimated this by placing the aperture at a
symmetric location on the other side of CoRoT-2, where the diffracted
light is contributed almost exclusively by CoRoT-2 itself.  Using that
diffracted fraction together with the flux ratio of M-dwarf to
CoRoT-2, we infer that the diffracted light from the M-dwarf
contributes 5.9\%, 5.0\%, and 8.3\% to CoRoT-2 at 3.6, 4.5 and
8.0\,$\mu$m, respectively. The eclipse photometry and derived
parameters for CoRoT-2 in our Figures and Tables have all been
corrected for this diffracted light.  \citet{gillon10} inferred 16.4\%
and 14.3\% at 4.5 and 8\,$\mu$m, respectively, but he used aperture
radii of 4.0 and 3.5 pixels, respectively, {\it vs.} 3.0 pixels in our
case.  

As a check, we repeated our diffracted light correction using
apertures having the same size as \citet{gillon10}. Because the
diffracted light is not uniform, the values do not simply scale as the
area of the aperture.  For the same apertures as \citet{gillon10}, we
obtain corrections of 14.2\% and 12.0\% at 4.5 and 8.0\,$\mu$m,
respectively, in reasonably good agreement with the independent
determination of \citet{gillon10}.  Uncertainty in the diffracted
light correction is not included in our eclipse amplitude error
estimates. Given our good agreement with the diffracted light
corrections of \citet{gillon10}, and given that we use smaller
photometric apertures than \citet{gillon10}, we conclude that
uncertainty in the diffracted light correction does not contribute
significantly to the errors on our measured eclipse depths.

The best-fit eclipse depths and errors are listed in Table~1, and the
central phases and errors are listed in Table~2.

\subsection{Error Estimation}

The ideal method to calculate errors would be to repeat all of the
observations and analysis, and compare the results from analyzing many
independent sets of observations.  This is obviously impractical, so
we mimic some key aspects of that ideal procedure. We generate fake
photometric datasets having the same properties as the real
photometry, and we repeat the entire iterative fitting process -
including intra-pixel corrections and ramp fitting - on each fake
dataset.  We calculate the standard deviation of the collection of
eclipse depths and central phases resulting from the repetitions of
the analysis on the fake data.

To generate each fake dataset, we subtract the best-fit eclipse curve
(plus baseline and intra-pixel decorrelation function) from the
original photometry to produce a set of photometric residuals. We
likewise produce a set of image position residuals by subtracting a
multi-point running average of the X and Y-pixel positions from each
individual (X,Y) position measurement. We permute all of the residuals
and add them back to the best-fit function (photometry) or running
average coordinate (position) to make an individual fake dataset.  We
permute the residuals using two methods, to make two distinct
collections of fake data.  The first permutation method scrambles the
residuals randomly, which is equivalent to the conventional bootstrap
Monte Carlo technique \citep{press}.  We generate $10^4$ bootstrap
datasets (trials) using this method, and calculate the standard
deviation of eclipse depth and central phase from the distributions of
these parameters over the $10^4$ trials.  These distributions are
close to Gaussian.

A second method to permute the residuals preserves their relative
order but shifts their initial phase; this is sometimes called the
`prayer-bead' method \citep{gillon09b}. In this case, the number of
trials equals the number of original photometry points.  This is 888
for CoRoT-1, and 13,545 for version 2 of the CoRoT-2 subarray
photometry.  These are adequate to define the distributions of eclipse
depth and phase.  The prayer-bead method is more sensitive to the
presence of red noise in the data. Nevertheless we find that the
distribution of eclipse depth remains consistent with a Gaussian, but
for CoRoT-1 the distribution of eclipse phase shows about $7\%$ of the
central phases lie below the $3\sigma$ point in the distribution.  We
attribute this to the presence of some red noise before mid-eclipse,
visible in the bottom panel of Figure~1.

For CoRoT-2, the distributions of eclipse depth and phase were close
to Gaussian, so errors from the prayer-bead method were quite close to
the values from the bootstrap method.  This indicates relatively
little red noise in the CoRoT-2 data (after we omitted the first 19
data points, as noted above).  For both CoRoT-1 and -2, we adopted the
greater of the bootstrap and prayer-bead errors for each
parameter. CoRoT-1 errors are uniformly larger than for CoRoT-2 because the star
is fainter and the red noise is slightly greater. Tables~1 \& 2 list the
errors on eclipse depth and central phase for all three eclipses, plus
our results from re-analysis of CoRoT-2 at 4.5 and 8\,$\mu$m (see
below).

\subsection{CoRoT-2 at 4.5\,and 8\,$\mu$m}

We check our methodology by analyzing archival Spitzer data for
CoRoT-2 at 4.5- and 8\,$\mu$m, for comparison to the results of
\citet{gillon10}. Our analysis at 4.5\,$\mu$m proceeds as described
above for CoRoT-1. At 8\,$\mu$m our eclipse fitting procedure uses a
`ramp' baseline \citep{deming06, knutson09} that is fit simultaneously
with the eclipse depth by linear regression.  The ramp is comprised of
a term linear in time, a term linear in the logarithm of time, with a
zero-point on the time axis as described by \citet{todorov}.  We also
find that the photometry exhibits a rather rapid decrease in flux
during the first 100 data points.  Investigating this, we find an
approximately 0.1-pixel change in the image Y-position during those
first 100 points.  This transient positional drift is not in sync with
the well known telescope pointing oscillation.  Although the pointing
oscillation has not (to our knowledge) been shown to affect 8\,$\mu$m
Spitzer photometry, the 0.1-pixel transient drift apparently does.  We
therefore include a Y-position term in the linear regression fit for
the eclipse depth.  Without this term, the eclipse depth would be
$0.42\%$, versus our result of $0.446\%$ (Table~1).

We also perform trial fits using the double-exponential ramp of
\citet{agol}.  These fits, like the log ramp discussed above, omit the
first 100 points and include a Y-position term.  The ramp observed in
the 8\,$\mu$m data (illustrated by \citealp{gillon10}) is very
shallow, and the scatter is relatively large compared to the
ramp-related flux change.  For this reason, we use a single
exponential ramp, not a double exponential ramp.  We experimented with
double-exponential fits, but our Levenberg-Marquardt fitting procedure
produced degeneracies when attempting to fit two exponentials to such
a shallow ramp. We believe that only one exponential is warranted in
this case. Moreoever, the best-fit exponential ramp is close to a
straight line, since the ramp curvature is minimal. As will become
apparent in Sec.~3.2, the 8\,$\mu$m eclipse depth of CoRoT-2 is
crucial to the interpretation of our results, so we will return to the
implications of fitting the exponential ramp during that discussion.

Our results for CoRoT-2 at 4.5 and 8\,$\mu$m are included in Tables~1
and~2.  The eclipse depth using the exponential ramp at 8\,$\mu$m is
included in Table~1, but the phase results for that ramp are the same
as the log ramp, and are not listed separately in Table~2. Overall, we find
excellent agreement with \citet{gillon10}.

\section{Results and Discussion}

\subsection{Orbital Phase}

For CoRoT-1, we compute the weighted average of the central eclipse
phase using both 3.6- and 4.5\,$\mu$m eclipses, adopting weights equal
to the inverse of the variance of each measurement. This yields a
central phase of $0.4994\pm0.0013$, and $|e\,cos(\omega)| < 0.006 $ to $3\sigma$.
Our limit indicates that the orbit is close to circular, but a small
non-zero eccentricity (such as we infer for CoRoT-2, see below) is not excluded.

For CoRoT-2, \citet{gillon10} found $e\,cos(\omega) = -0.00291\pm
0.00063$.  Our result for the 3.6\,$\mu$m eclipse (central phase at
$0.4994\pm0.0007$) is displaced in the same direction as
\citet{gillon10} infer, but with insufficient precision to confirm or
reject the \citet{gillon10} claim.  Combining our 3.6\,$\mu$m result
with the eclipses analyzed by \citet{gillon10} could increase the
significance of the total result.  For maximum consistency, we
re-analyzed the 4.5- and 8\,$\mu$m eclipse data, as described above.
We verified that our adopted transit ephemeris (see Table~2) should
not be a significant source of error when propagated to the eclipse
times. Weighting each eclipse phase (3.6, 4.5 and 8, see Table~2) by
the inverse of its variance yields an average central phase of
$0.49809\pm0.00028$.  Including the 28~seconds for light to cross the
planetary orbit, we expect to find the eclipse at phase $0.500019$ if
the orbit is circular. Hence, we derive $e\,cos(\omega) =
-0.0030\pm0.0004$.  The excellent agreement with \citet{gillon10} is
in part because we are analyzing much of the same data. However, the
result is heavily weighted by the single eclipse at 4.5\,$\mu$m, which
is a reason to be cautious concerning a claim of non-zero
eccentricity.  Nevertheless, at face value we are able to reproduce
the result of \citet{gillon10} using an independent analysis, and
improve the precision slightly.  \citet{gillon10} point out that a
non-zero eccentricity does not require an additional planet in the
system, since incomplete two-body tidal circularization is a plausible
alternative for this system.

\subsection{Atmospheric Temperature Structure}

Our results for both planets are summarized in Figure~4, which shows
all available eclipse data in comparison to various models.  The
caption of Figure~4 gives reduced $\chi^2$ values for the comparison
between each model and the eclipse data.  Since Figure~4 is a
comparison of the data to model predictions, not a fit involving
adjustable parameters, we take the degrees of freedom to equal the
number of data points when calculating the reduced $\chi^2$.

The model comparison for CoRoT-1 (top panels of Figure~4) suggest an
inverted atmospheric temperature structure. The best overall account
of the data is actually produced using a $T=2460K$ blackbody spectrum
(\citealp{rogers}, green line, see reduced $\chi^2$ values in Figure~4
caption). However, this likely indicates the presence of a high
altitude absorbing layer, and such layers are implicated in driving
the inversion phenomenon \citep{burrows07, knutson08}. The nature of
the absorber is the subject of current debate \citep{fortney3,
spiegel}. The conventional model (black line, \citealp{burrows08})
shows significant absorption due to the CO bandhead that occurs near
4.7\,$\mu$m, and the Spitzer data show no sign of being affected by
this feature.  An inverted model using TiO absorption (blue line)
shows much better agreement with the data than the non-inverted model,
but does not account particularly well for the ground-based
(2\,$\mu$m) measurements.  An atmosphere with a nearly isothermal
region over extended heights will produce a blackbody-like spectrum,
and can be regarded as a special case of an inverted temperature
structure.  The inverted and blackbody model for CoRoT-1 both give
good agreement with the Spitzer data, as well as the CoRoT optical
eclipse measurements (\citealp{snellen09, alonso09b}). This indicates
that the optical emission is predominately thermal in origin.  The
models that account for our Spitzer data, when compared to the optical
eclipses (Figure~4), leave little room for a reflected light
component.  Based on the models of \citet{sws}, a geometric albedo
near unity would produce a reflected light eclipse depth of
approximately 520 ppm, whereas the difference between the CoRoT-1
observations \citep{snellen09} and the inverted model (blue curve on
Figure~4) is 84 and 21 ppm at 0.6 and 0.71\,$\mu$m,
respectively. Also, \citet{cowan} inferred a Bond albedo of $<10$\% for
CoRoT-1. Our results therefore support the conclusion of
\citet{snellen09} and \citet{cowan} that CoRoT-1 is a dark planet.

CoRoT-2 (bottom panels of Figure~4) is more complex than CoRoT-1.  A
conventional model (black line, \citealp{burrows08}) produces
excellent agreement with all of the data except for the 4.5\,$\mu$m
point, where the disagreement is substantial. Since the 4.5- to
3.6\,$\mu$m contrast ratio is even greater than for CoRoT-1, a
temperature inversion is suggested.  But inverted models do not
reproduce the 8\,$\mu$m contrast and, based on the reduced $\chi^2$
values (Figure~4 caption), no model gives a reasonable account of the total
data. Both the 4.5 and 8\,$\mu$m observed values are in good agreement
between our analysis and \citet{gillon10}, so the problem does not
seem to lie with the observations.  We first mention some caveats, and
then we suggest two hypotheses to account for the contrast values of
this unusual planet.

One caveat that applies to CoRoT-2 is the fact that the star is active
\citep{alonso08}. However, because the planet passes behind the star
during eclipse, there is no time-variable blocking of active regions
on the stellar disk.  The primary consequence of stellar activity is
the photometric variation of the star itself. This variation can
manifest itself in two ways.  First, stellar variations can appear
directly in the eclipse curve.  The dominant stellar variation will be
due to rotational modulation of active regions, with a 4.5-day period
\citep{lanza}.  This time scale is more than an order of magnitude
longer than the 2.2-hour eclipse duration.  Although rotation of
active regions can still affect eclipse data (e.g., by perturbing the
photometric baseline) we do not discern any indications of it, so we
interpret our data at face value. The second way in which stellar
variations can affect eclipse depth is through the normalization.
When the star is fainter, the disappearance of the planet during
eclipse translates to a larger fraction of the stellar flux.  This
effect can alter eclipse depths on long time scales.  However, the 4.5
and 8\,$\mu$m observations made by \citet{gillon10} were simultaneous,
so long-term stellar variability cannot be a factor in the puzzling
spectrum of CoRoT-2.

A final caveat concerns the ramp effect for CoRoT-2 at 8\,$\mu$m.  We
find that fitting the exponential model of \citet{agol} increases the
eclipse depth to 0.51\% (Table~1).  However, this does not alter the
situation concerning the interpretation of the CoRoT-2 results, so we
now discuss two hypotheses to account for the totality of the CoRoT-2
data as summarized in Tables~1~\&~2.

\subsection{Possible Mass Loss for CoRoT-2}

Our first hypothesis for CoRoT-2 is that the planetary atmosphere is
well described by a conventional (non-inverted) model, but the
4.5\,$\mu$m eclipse appears anomalously deep because it contains
carbon monoxide emission lines due to mass loss.  We find that a
conventional model lacking CO absorption (see Figure~4) does not
increase the contrast sufficiently in the 4.5\,$\mu$m band to account
for the data - the reduced $\chi^2$ is 13.5 (Figure~4). Actual
emission from mass loss would be required.  Mass loss for close-in
giant exoplanets can occur via tidal stripping \citep{li}, and also
via energy deposition from stellar UV flux. The latter process is
particularly important for planets orbiting young, UV-bright stars
\citep{baraffe, hubbard}.  CoRoT-2 orbits very close-in, where the
tidal force is strong (0.026 AU, \citealp{barge}). Moreover, the star
is young and active \citep{bouchy}, possibly as young as 30 Ma
\citep{guillot}. Hence both mass loss mechanisms are potentially
important for this planet.

\citet{li} have predicted significant CO emission in the $\Delta{V}=2$
overtone bands near 2.29\,$\mu$m, due to tidally-stripped mass loss
from WASP-12. This mass should also emit in the CO $\Delta{V}=1$
bands, which are intrinsically stronger than the overtone bands, and
arise from upper levels that are easier to excite.  Emission from the
$\Delta{V}=1$ bands will fall within the 4.5\,$\mu$m bandpass,
increasing the eclipse depth.  Tidal-induced mass loss is at least
qualitatively consistent with the apparent non-zero eccentricity of
the orbit. However, recent results show that the orbit of WASP-12b is
likely to be more circular than \citet{li} suppose \citep{campo,
husnoo}.  The evidence for non-circularity is better in the CoRoT-2
case than for WASP-12, so we explore whether a mass loss and CO
emission scenario might be profitably applied to CoRoT-2.

We calculate what mass loss rate is required to increase the
4.5\,$\mu$m contrast sufficiently over the conventional model to
account for the observed eclipse depth.  We compare the requisite mass
loss rate with model calculations for both tidal-stripping, and
evaporation by stellar UV flux. If the required mass loss rate is (for
example) so large that the planet would disappear within an
unacceptably small time scale, then we could discard the mass loss
hypothesis.

Prior to calculating the mass loss required to account for the
4.5\,$\mu$m eclipse, we mention a potentially serious problem with
this hypothesis.  This problem derives from the eclipse curve itself.
In a variant of our bootstrap error analysis, we allowed the ingress
and egress times of the eclipse to vary.  We implemented variations in
ingress/egress time by applying linear transformations to the time
axis prior to second contact, and subsequent to third contact.  We
find that the 1$\sigma$ precision of the observed ingress/egress time
is about 10\%.  This implies that the radius of any CO-emitting volume
cannot be more than about 30\% larger (3$\sigma$ limit) than the
radius of the planet.  Given the requisite mass loss rate (see below),
we calculated a synthetic spectrum for the resultant CO column density
of $10^{19}$ cm$^{-2}$, adopting excitation temperatures from 3000K to
15,000K.  Many individual lines in this spectrum are optically thick,
and attain intensities closely equal to the Planck function at the
excitation temperature.  However, the line density in the 4.5\,$\mu$m
Spitzer bandpass is insufficient to produce the required eclipse flux
unless the excitation temperature exceeds 15,000K.  Since CO is
primarily dissociated at such temperatures, we cannot easily match the
required eclipse flux using such a compact source of CO
emission. Nevertheless, the details of mass loss in the Roche lobe and
through the inner Lagrangian point are not completely understood, so
we present our calculation of the mass loss required to account for
the 4.5\,$\mu$m eclipse depth.

Let the continuum flux from the star, integrated over the 4.5\,$\mu$m
band be denoted $F_s$, in ergs cm$^{-2}$ sec$^{-1}$.  Let the flux
from the hypothetical CO cloud be denoted $F_{CO}$ in the same
units. Then the excess over the standard model atmosphere for the
planet (Figure~4) requires:

\begin{equation}
F_{CO} \approx 0.005F_s  
\end{equation}

A Phoenix model atmosphere for the star \citep{phoenix}, integrated
over the 4.5\,$\mu$m bandpass, gives the same flux as blackbody having
$T=5237$K, so $F_s = {\Delta\nu}\Omega_{s}B_{\nu}$, where $B_{\nu} =
5.17 \times 10^{-6}$ is the Planck function (in cgs units) at 5237K,
$\Omega_{s}$ is the solid angle of the star as seen from Spitzer, and
$\Delta\nu$ is the bandwidth of the 4.5\,$\mu$m band in Hz.  We also
have $F_{CO} = L/(4\pi d^2)$, where $L$ is the luminosity of the
CO-emitting cloud within the 4.5\,$\mu$m band (ergs sec$^{-1}$), and
$d$ is the distance to the system.  The solid angle $\Omega_{s} = \pi
R^2/d^2$, where $R$ is the radius of the star.  We substitute for
$d^2$ in the expression for $F_{CO}$, and then (1) becomes:

\begin{equation}
L \approx 0.2\Delta\nu B_{\nu} R^2 \approx 6.2 \times 10^{28} ergs~sec^{-1}
\end{equation}

The number of CO molecules required to produce this luminosity depends
on their excitation state and on the Einstein-A values for the
emission.  We first adopt a thermal distribution at $T=3000$K for the CO
vibrational levels, and we use the rotationless Einstein A-values
$A_{ji}$ for $\Delta V =1$ from \citet{okada}.  Summing over the
vibrational levels, we find that the effective emitting rate is $28$
sec$^{-1}$ per CO molecule.  Since $h\nu \approx 4.42 \times
10^{-13}$, $L \approx 1.4 \times 10^{41}$ photons sec$^{-1}$.  This
requires $4.9 \times 10^{39}$ CO molecules in the emitting volume.
Adopting a solar carbon abundance ($10^{-3.5}$), and stipulating that
all of the carbon appears in CO, the total mass in the emitting volume
is approximately $1.5 \times 10^{-11}$ Jupiters.

To determine a mass loss rate from the total mass in the emitting
volume, we must estimate the transit time of CO molecules.  This has
been discussed by \citet{li}, who conclude that mass flows through the
Roche lobe at the sound speed, and forms a disk around the star.  Most
of that disk emission will not be modulated by the secondary eclipse,
so our observations refer only to the mass flowing out of the Roche
lobe itself. The relevant time is therefore the Roche lobe radius
$a(M_p/3M_s)^{1/3}$ divided by the sound speed $(\gamma
P/\rho)^{1/2}$.  We calculate a Roche lobe radius for CoRoT-2 of $4.3
\times 10^5$ km, and a sound speed of $4.5$ km sec$^{-1}$.  These
values yield a mass loss rate of $\sim 5 \times 10^{-9} M_J$ per year.
This value is in close accord with a minimum value for WASP-12,
calculated by \citet{lai}.  It is also a reasonable value for a giant
planet close-in to a young active star \citep{hubbard}.

The greatest uncertainty in the above calculation is the excitation
state of the CO molecules.  Because the population of the vibrational
levels varies exponentially with vibrational temperature, the
effective emitting rate could vary by orders of magnitude and still be
consistent with our ignorance.  If CO lost from the planet is
vibrationally cold (T=300K, for example), as will tend to happen in
the absence of collisional excitation, then the effective emission
rate drops by over 4 orders of magnitude, and the required mass loss
rate increases by that factor, and becomes unacceptably large. Indeed,
in the arguably applicable limit of no collisional excitation, each CO
molecule would emit approximately one photon as it expanded from the
planetary atmosphere through the Roche lobe. That limit would require a
mass loss rate as high as $10^{-2} M_J$ per year, which is
unacceptably high.

Although the requisite mass loss rate is within the range for
tidal-stripping and UV-energy deposition models, we conclude that this
CO-emission hypothesis is an unlikely interpretation of the Spitzer
data, due to the difficulty with the ingress/egress time and the
necessity of maintaining collisional excitation. However, it cannot be
absolutely ruled out without more detailed models as well as observed
high resolution spectroscopy of the system. If this hypothesis could
be confirmed, the consequent lack of an atmospheric temperature
inversion for this planet - orbiting an active star - would be
consistent with the emerging anti-correlation between the presence of
inversions and stellar activity levels \citep{knutson10}.

\subsection{An Inverted Atmosphere Variant for CoRoT-2}

A second hypothesis to account for CoRoT-2b is a variant of an
inverted atmospheric structure. The 8\,$\mu$m radiation may
hypothetically emerge from deeper and cooler atmospheric layers,
whereas the shorter wavelengths are formed in a high altitude layer
that is heated by absorption.  Absorption in a high altitude layer has
been implicated \citep{burrows07} as driving atmospheric temperature
inversions, by absorbing stellar irradiance and heating the planetary
atmosphere at altitude. Radiative equilibrium of a high altitude
absorbing layer that is optically thick in the optical and near-IR
could potentially shield lower levels of the atmosphere from radiative
heating. A high altitude layer would re-emit both to space and to
lower levels of the atmosphere, but the net downward flux would be
reduced by upward emission to space.  If the opacity of the absorbing
layer is high in the optical and near-IR ($\lambda < 5\mu$m), eclipse
observations at those wavelengths may sense only the absorbing layer,
whereas longer wavelengths (e.g., 8\,$\mu$m) may penetrate and sense
the cooler lower atmosphere. 

Recently, \citet{guillot} have concluded that the IR opacity of
CoRoT-2's atmosphere is greater than normal.  We are here
hypothesizing exactly the opposite of that conclusion, but based in
part on the additional 3.6\,$\mu$m eclipse result that was not available to
\citet{guillot}.

One immediate problem with this hypothesis is that 8\,$\mu$m radiation
is not believed to be formed any deeper than the shorter wavelength
IRAC bands \citep{burrows07}. Hence some additional source of short
wavelength opacity is required. Scattering by micron-sized haze
particles or aerosols is a potential source of the required opacity if
such particles can be lofted and maintained at high altitudes. Haze
due to smaller particles at high altitudes has been inferred for other
planets \citep{pont}.  However, several caveats should be cited
with regard to this hypothesis.  First, most scattering opacities from
small particles have a very broad dependence on wavelength, whereas a
sharper long-wavelength cutoff might be required.  If the extra
opacity is from absorption (as opposed to scattering) then it might
perturb the atmospheric temperature gradient so that the cooler lower
atmosphere we envision might not exist.   

This hypothesis of a heated high altitude layer and a cooler lower
atmosphere brings to mind the situation with respect to the global
energy budget of HD\,189733b. \citet{barman} pointed out that the
efficiency of zonal heat redistribution can be highly depth
dependent. Deeper layers can redistribute heat more efficiently
because their radiative time constant \citep{Iro} is comparable to or
exceeds the time for advection of heat by zonal winds. In that case
the lower atmosphere responds primarily to the day-night average
irradiation, whereas the upper atmosphere comes to radiative
equilibrium with day-side irradiation on a short time scale. If
8\,$\mu$m radiation from CoRoT-2 arises from deeper layers, then this
effect can in principle act to reinforce the presence of a temperature
inversion.

If this second hypothesis is correct, then high opacity at optical
and near-infrared wavelengths could produce a blackbody spectrum at
these wavelengths.  An 1866K blackbody (green line on Figure~4)
produces a reasonable agreement with the 3.6 and 4.5\,$\mu$m data, but
is below the optical CoRoT measurements \citep{alonso09a, snellen10}.
\citet{cowan} invoked a simple analytic model of the published photometric
observations of close-in exoplanets, and inferred $T=1866$K and a Bond albedo of
$16\%\pm7\%$ for CoRoT-2b. This is qualitatively consistent with our
second hypothesis for this planet.  Ground-based JHK eclipse
measurements of this unusual planet would be very useful in defining
the blackbody shape and temperature of the near-infrared spectrum.

\acknowledgements This work is based on observations made with the
{\it Spitzer Space Telescope}, which is operated by the Jet Propulsion
Laboratory, California Institute of Technology, under a contract with
NASA.  Support for this work was provided by NASA. Eric Agol
acknowledges support under NSF CAREER grant no. 0645416.  Adam Burrows
was supported by NASA grant NNX07AG80G and under JPL/Spitzer
Agreements 1328092, 1348668, and 1312647.  He is also pleased to note
that part of this work was performed while in residence at the Kavli
Institute for Theoretical Physics, funded by the NSF through grant
no. PHY05-51164. We thank Dr. Rory Barnes for informative
conversations regarding the tidal evolution of CoRoT-2, and an
anonymous referee for a very thorough review that improved this paper
significantly.
\\

\appendix
Because our results are among the first for exoplanets using Warm
Spitzer, we take this opportunity to comment on the photometric
quality of the Warm mission exoplanet data.  The loss of cryogen has
increased the operating temperature of the InSb detectors from 15K (cryogenic) to
29K (warm), and that has altered some characteristics of the detectors.  For
example, the `column pull down' effect has become more prominent.  Bright stars
cause the signal levels to drop for all pixels in the column they
overlap.

None of our target stars happen to lie on columns that are noticeably
affected by pull-down.  Our photometry code calculates the theoretical
limiting signal-to-noise ratio based on the Poisson statistics of the
total number of electrons recorded from the star, and we include a
read noise of 10 electrons for each pixel within the numerical
aperture.  After fitting the photometric time series to remove the
intra-pixel variations and the eclipse, we calculate the scatter of
the residuals and compare this to the theoretical limiting noise.  For
CoRoT-1 at 3.6 and 4.5\,$\mu$m we achieve 87\% and 92\% of the
theoretical signal-to-noise, respectively.  However, this seemingly
excellent performance may be mis-leading because these are relatively
faint stars, where the stellar photon noise is high and will tend to
dominate instrumental noise.  A more sensitive test for possible
instrumental red noise is to calculate the reduced $\chi^2$ of the
binned data, after removing the best fit eclipse (bottom panels of
Figs.~1-3). We base the predicted error of each bin (error bars on the
figures) on the {\it observed} scatter of the unbinned points,
reduced by the square-root of the number of points in each bin
(typically, 9).  On this basis, the reduced $\chi^2$ values are 1.10
and 1.31 for CoRoT-1 at 3.6 and 4.5\,$\mu$m, respectively.  This
indicates that a small amount of red noise occurs for time scales
longer than about 5 minutes.

In the case of CoRoT-2, the only binning we used was the averaging
over 64-frames in each data cube.  Measuring the observed scatter
after removing the fit, we find a ratio of 83\% when using all
individual frames of each 63-frame data cube, but this reduces to 75\%
of the theoretical signal-to-noise when we average the frames in each
data cube before fitting the eclipse.  Like CoRoT-1, this indicates the
presence of a small amount of red noise.

We are interested in whether the column pull-down effect causes
enhanced noise for stars that lie on affected columns.  Unfortunately,
there are no suitably bright stars that overlie pulled-down columns in
our CoRoT data,  nor did we find any optimal test stars in several
other Warm Spitzer data sets that we examined.  The best test star we
located was HD\,189314, lying in the Kepler field (D. Charbonneau, PID
60028).  This relatively bright star (K=9.3) is above the 1\%
nonlinearity limit for the 12-sec exposures we examined.  Because
pointing jitter moves the star toward and away from pixel-center, it
modulates the nonlinearity effect simultaneously with the intrapixel effect. We
were unable to effectively decorrelate these mixed instrumental
effects.  However, we were able to evaluate the point-to-point scatter in
the photometry, by removing a smoothing function (high-pass
filtering).  We find that the point-to-point scatter in the photometry
achieves 76\% of the theoretical signal-to-noise.  We tentatively
conclude that the column pull-down effect does not add short-term
noise to Warm Spitzer photometry, even for stars overlying affected
columns.  We are unable to evaluate whether it causes increased red
noise, but we anticipate that this will become clear as additional
Warm Spitzer observations are accumulated.

Finally, we draw attention to another important difference between the
cryogenic and Warm missions. With cryogenic data, we sometimes
evaluated the background for subarray photometry by considering a
median over all pixels in a data cube (fitting to a distribution), and
using this single best-fit background value for each of the 64-frames
in the data cube.  This had the advantage that the larger number of
pixels over the entire data cube resulted in a more precisely
determined value, but it was premised on the background being constant
within each data cube.  We find that this premise is no longer
accurate for the Warm mission: the background value varies
significantly from frame to frame within a subarray data cube.  (The
background is probably not due to impinging IR radiation, but is more
likely to be electronic in nature.) The statistical penalty of having
fewer pixels available when measuring the background in individual
frames is offset by the necessity of following these frame-to-frame
variations.  

The background variations are illustrated in Figure~5, where we show
the 3.6\,$\mu$m background per frame as a function of the frame number
within a data cube, and compare the cryogenic mission (represented by
HD\,189733) to the Warm mission (represented by CoRoT-2).  Note that the
58th frame continues to exhibit a higher background value in the Warm
mission, as it did in the cryogenic mission \citep{harrington, agol}.  We
find, in agreement with \citet{agol}, that the photometry from the 58th
frame is well-behaved if the higher background is accounted
for.  Because the Warm mission will inevitably observe fainter
exoplanet host stars than during the cryogenic mission, accurate
background subtraction becomes a high priority. Our 3.6\,$\mu$m
photometry for CoRoT-2 used the `per-frame' method that we now find to
be necessary, and achieved the 83\% of theoretical signal-to-noise as
described above.



{\it Facilities:} \facility{Spitzer}.




\clearpage



\begin{figure}
\epsscale{.60}
\plotone{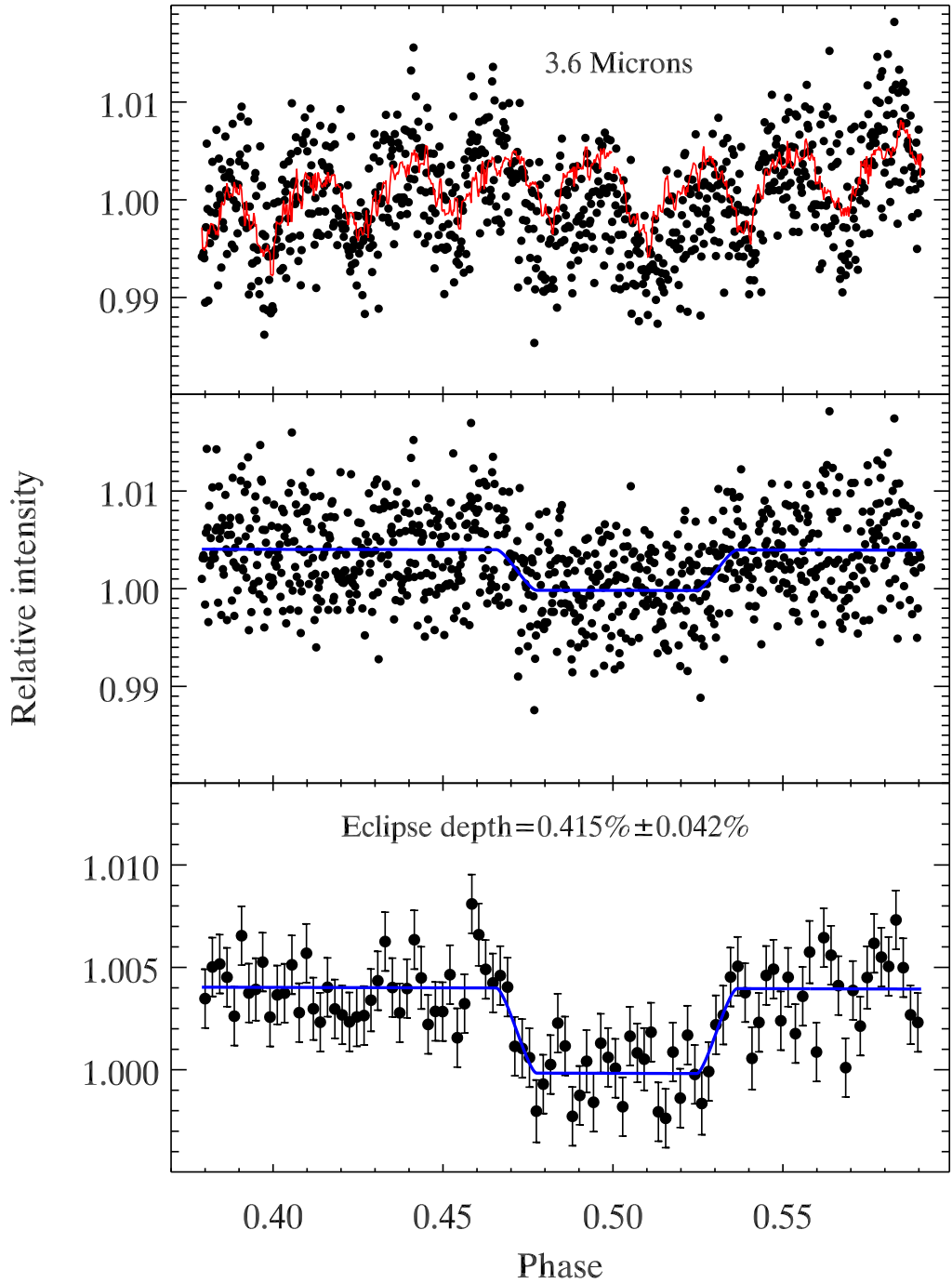}
\vspace{0.8in}
\caption{{\it Upper Panel:} Photometry of CoRoT-1, 
{\it vs.} orbital phase, at 3.6\,$\mu$m (points), with the decorrelation function
overplotted (red line).  {\it Middle Panel:} Photometry after
correction with the decorrelation function, with the best-fit eclipse
curve overlaid (blue line).  {\it Bottom Panel:} Decorrelated photometry binned to
a resolution of approximately 0.002 in orbital phase (100 bins), with
the best fit eclipse curve overlaid (blue line).  The error bars are based on the
scatter of individual points within each bin.  The best-fit central phase is
$0.5012\pm0.0024$.
\label{fig1}}
\end{figure}

\clearpage

\begin{figure}
\epsscale{.60}
\plotone{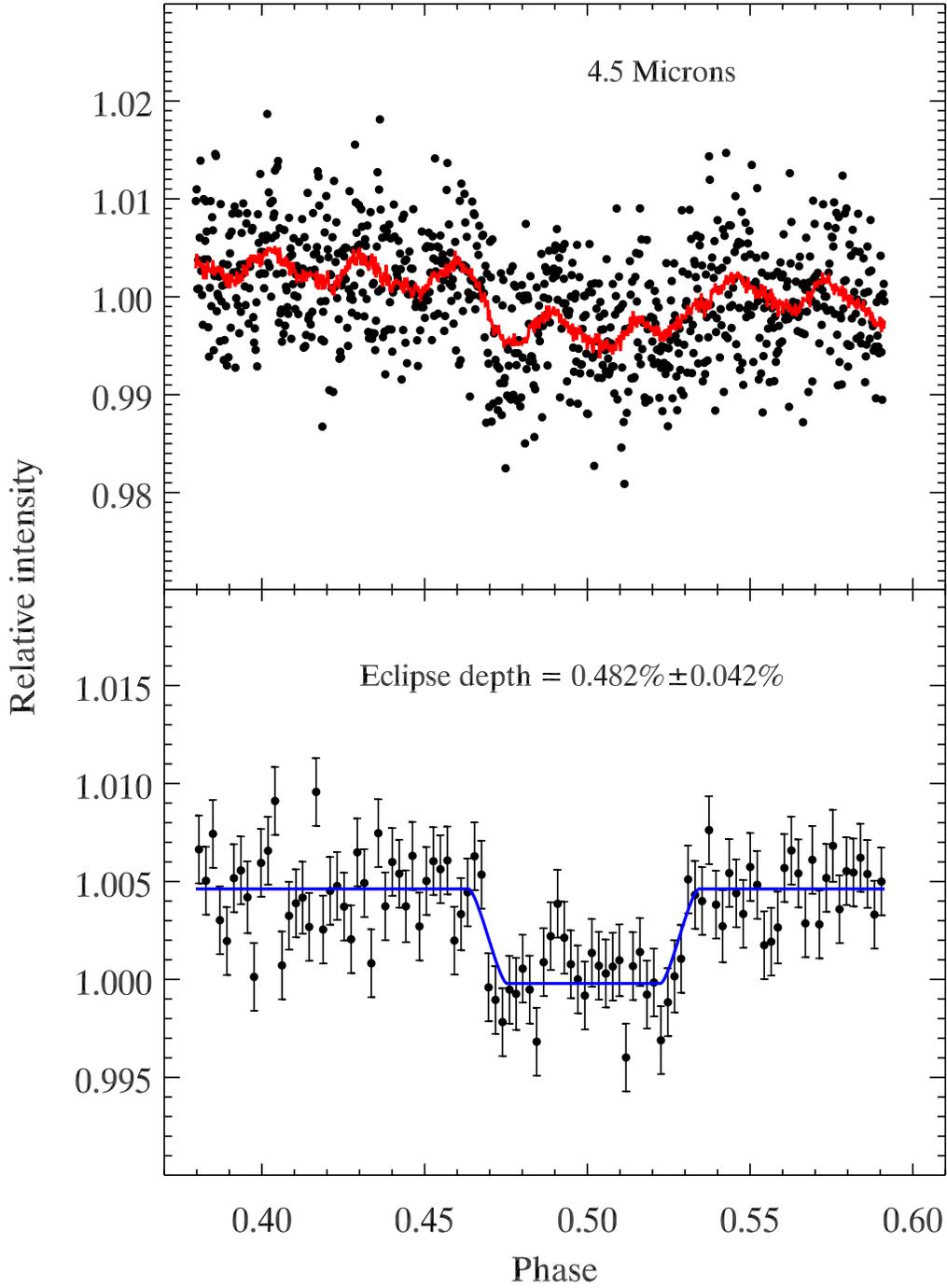}
\vspace{0.8in}
\caption{{\it Upper Panel:} Photometry of CoRoT-1, {\it vs.} orbital
phase, at 4.5\,$\mu$m (points), with the decorrelation function
overplotted (red line). {\it Bottom Panel:} Photometry with the
decorrelation function removed, and binned to a resolution of
approximately 0.002 in orbital phase (100 bins), with the best fit
eclipse curve overlaid (blue line).  The error bars are based on the
scatter of individual points within each bin.  The best-fit central phase is
$0.4992\pm0.0014$.
\label{fig2}}
\end{figure}

\clearpage

\begin{figure}
\epsscale{.60}
\plotone{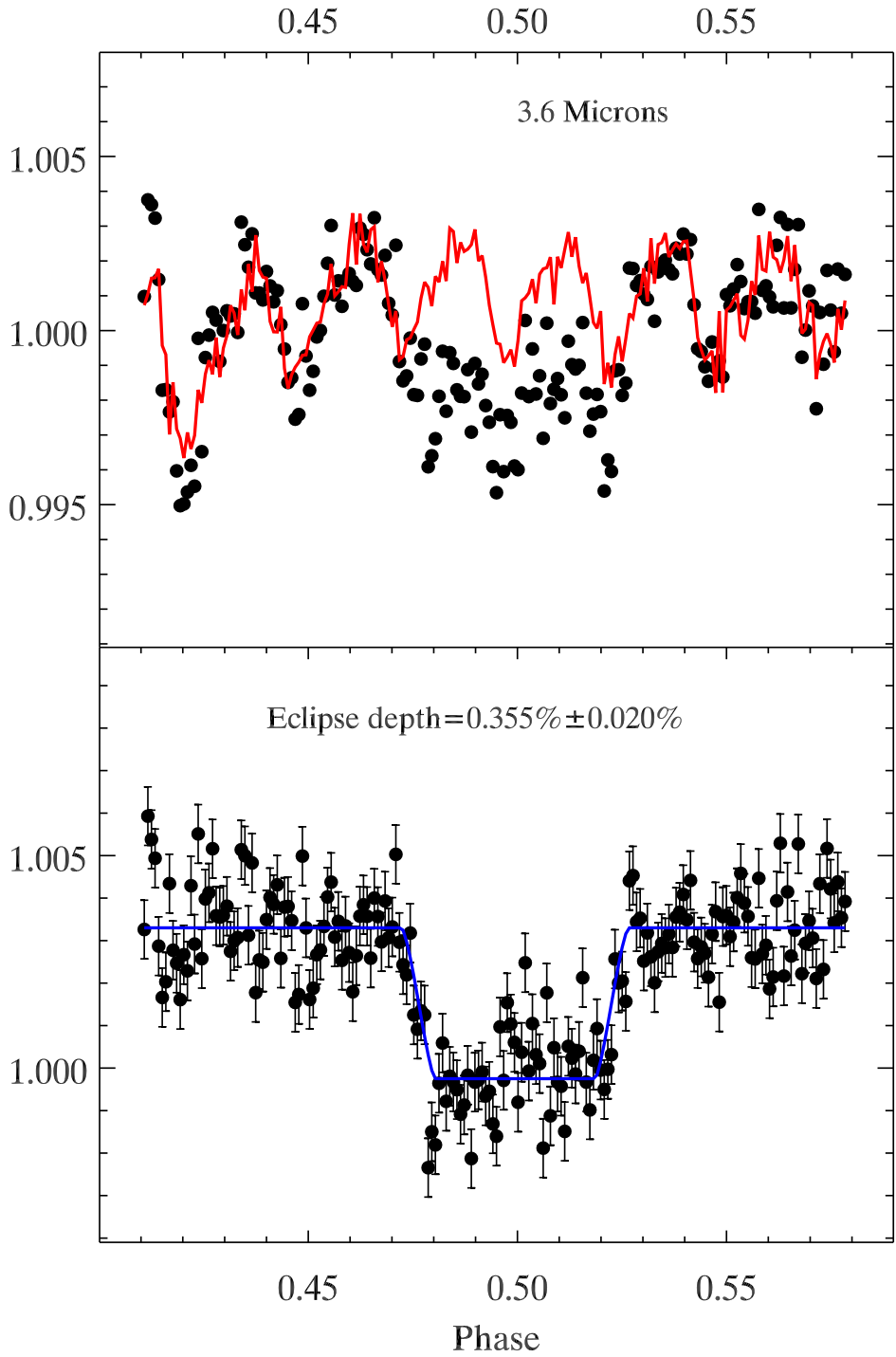}
\vspace{0.8in}
\caption{{\it Upper Panel:} Photometry of CoRoT-2, {\it vs.} orbital
phase, at 3.6\,$\mu$m (points), with the decorrelation function
overplotted (red line). Each point is the average of 63 temporal
frames in a data cube of $32 \times 32$ pixels times 64 temporal
frames (dropping the first). {\it Bottom Panel:} Photometry with the
decorrelation function removed, with the best fit eclipse curve
overlaid (blue line). The error bars are the theoretical limit based
on the photon and read noise.  The best-fit central phase is
$0.4994\pm0.0007$.
\label{fig3}}
\end{figure}

\clearpage

\begin{figure}
\epsscale{.60}
\plotone{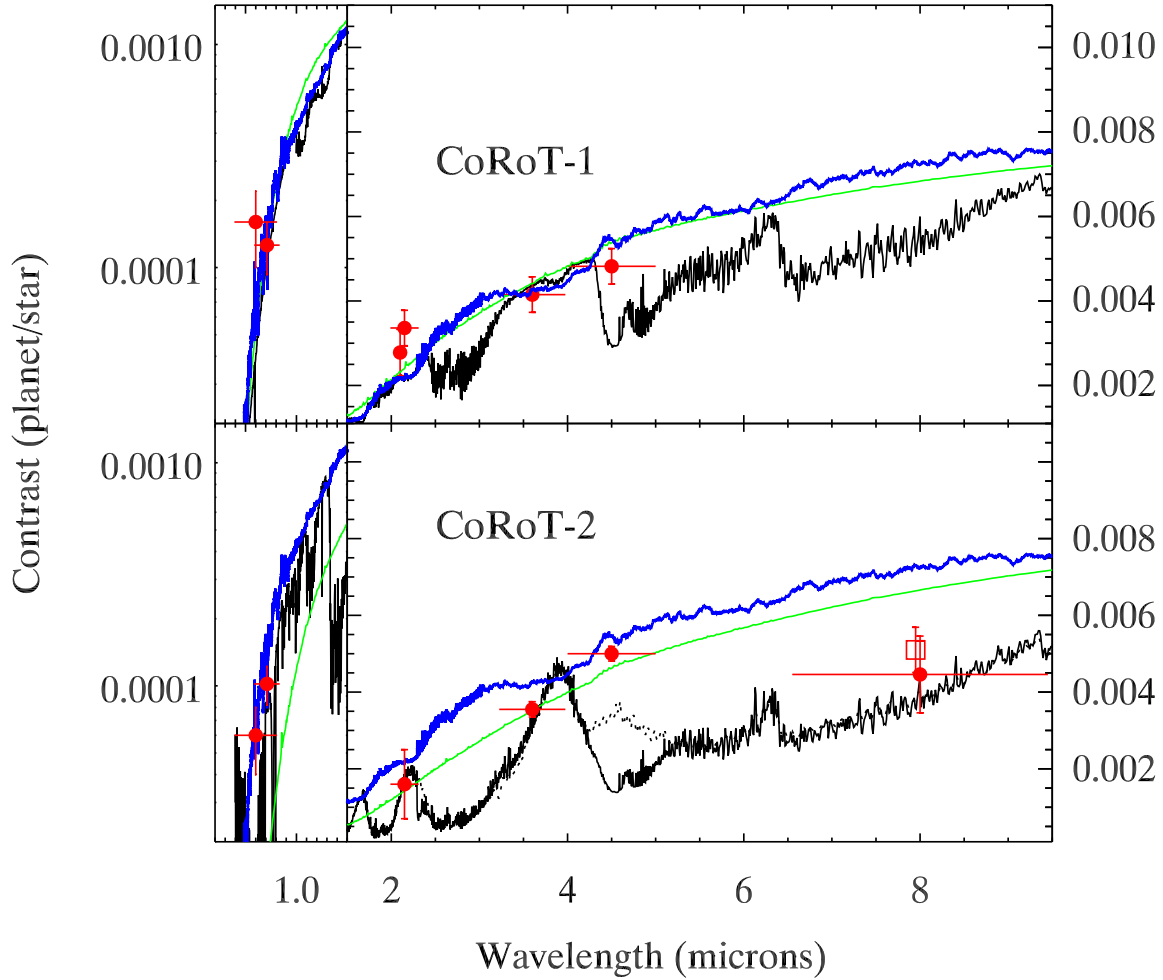}
\vspace{0.8in}
\renewcommand{\baselinestretch}{0.95}
\caption{\small Planet to star contrast ratios for CoRoT-1 and CoRoT-2
versus wavelength, from Table 1. The short wavelength data are on left
panels (contrast on log scale) and longer wavelength data on the right
(contrast on linear scale). Data from CoRoT, ground-based at
2\,$\mu$m, and Spitzer are all plotted with red points. Error bars on
the abscissa give the half-intensity wavelength limits of the
bandpasses. For CoRoT-2 we plot our re-analysis of the
\citet{gillon10} data at 4.5 and 8\,$\mu$m, but the original
\citet{gillon10} values are similar. The square point at 8.0\,$\mu$m
is the eclipse depth using the exponential ramp (see table~1). The
black curves are non-inverted Burrows models having 30\%
redistribution of stellar irradiance to the night side, with no extra
absorbing layers at high altitude. For CoRoT-2, the black dotted
portion near 4.5\,$\mu$m is the same Burrows model, only lacking CO
absorption. The blue lines are inverted models from Fortney and
collaborators \citep{fortney1, fortney2, fortney3} having TiO
absorption, and no re-distribution of stellar irradiance. The green
lines are blackbodies having $T=2460$K (CoRoT-1, \citealp{rogers}) and
$T=1866$K (CoRoT-2, \citealp{cowan}). The reduced $\chi^2$ values for
the CoRoT-1 data as compared to the different models are: conventional
model (black line) = 12.6, inverted model (blue line) = 2.4, blackbody
(green line) = 1.9.  For CoRoT-2, the reduced $\chi^2$ values for
those models are 61.4, 30.4, and 12.5, respectively.  (These values use the log
ramp point at 8\,$\mu$m, not the exponential ramp.)  The reduced
$\chi^2$ value for CoRoT-2 compared to the non-inverted model without
CO absorption (dotted portion) is 13.5. \normalsize
\label{fig4}}
\normalsize
\renewcommand{\baselinestretch}{1.0}
\end{figure}

\clearpage

\begin{figure}
\epsscale{.70}
\plotone{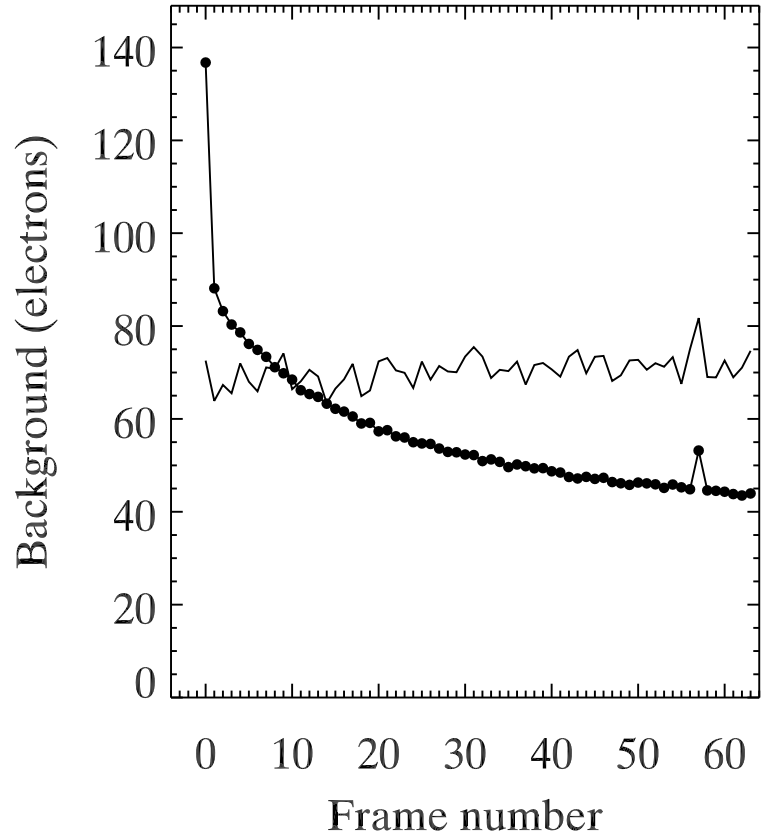}
\vspace{0.8in}
\caption{Number of electrons per pixel in the background of CoRoT-2 at
  3.6\,$\mu$m (points with line connecting), shown as a function of
  the frame number in each 64-frame subarray data cube observed using
  Warm Spitzer.  These results are averaged over all 215 data cubes
  that were acquired, and the exposure time per frame was 2~seconds.
  The line without points shows the background for subarray photometry
  of HD\,189733, using observations acquired during the cryogenic
  mission \citep{charb08}.  Since background contains both real
  infrared radiation as well as electronic effects, it is not
  proportional to exposure time.  The short-exposure (0.1-sec)
  HD\,189733 observations were scaled upward by an arbitrary factor
  for this plot.
\label{fig5}}
\end{figure}

\clearpage

\begin{table}
\begin{center}
\caption{Summary of Secondary Eclipse Measurements for CoRoT-1 and CoRoT-2 \label{tbl-1}}
\begin{tabular}{clll}
\tableline\tableline
Planet & Wavelength & Eclipse Depth & Reference\\
\tableline
 CoRoT-1 & 0.60(0.42) $\mu$m   & $0.016\%\pm0.006\%$   & Alonso et al.(2009b) \\
 --      & 0.71(0.25)          & $0.0126\%\pm0.0033\%$ & Snellen et al.(2009) \\
 --      & 2.10(0.02)          & $0.278\%^{+0.043\%}_{-0.066\%}$ & Gillon et al.(2009)   \\
 --      & 2.15(0.32)          & $0.336\%\pm0.042\%$ &  Rogers et al.(2010) \\
 --      & 3.6(0.75)          &  $0.415\%\pm0.042\%$  &  {\bf This paper} \\
 --      & 4.5(1.0)          &   $0.482\%\pm0.042\%$  &  {\bf This paper} \\
 CoRoT-2 & 0.60(0.42) $\mu$m   & $0.006\%\pm0.002\%$   & Alonso et al.(2009a) \\
 --      & 0.71(0.25)          & $0.0102\%\pm0.002\%$ & Snellen et al.(2010) \\
 --      & 2.15 (0.32)         & $0.16\%\pm0.09\%$  &  Alonso et al.(2010) \\
 --      & 3.6(0.75)           & $0.355\%\pm0.020\%$ & {\bf This paper}   \\
 --      & 4.5(1.0)          & $0.510\%\pm0.042\%$ &  Gillon et al.(2010) \\
 --      & 4.5(1.0)          & $0.500\%\pm0.020\%$ &  {\bf This paper} \\
 --      & 8.0(2.9)          &  $0.41\%\pm0.11\%$  &  Gillon et al.(2010) \\
 --      & 8.0(2.9)          &  $0.446\%\pm0.10\%$  &  {\bf This paper - log ramp}  \\
 --      & 8.0(2.9)          &  $0.510\%\pm0.059\%$  &  {\bf This paper - exponential ramp}  \\
\tableline
\end{tabular}
\end{center}
\end{table}

\begin{table}
\begin{center}
\caption{Eclipse Central Times and Phase for CoRoT-1 and CoRoT-2.\label{tbl-2}}
\begin{tabular}{clll}
\tableline\tableline
Planet & Wavelength & HJD  & Phase \\
\tableline
 CoRoT-1 & 3.6 $\mu$m   & $2455162.1643\pm0.0036$ &   $0.5012\pm0.0024$  \\
         & 4.5          & $2455159.1433\pm0.0021$ &   $0.4992\pm0.0014$   \\
 CoRoT-2 & 3.6          & $2455160.4496\pm0.0012$ &   $0.4994\pm0.0007$  \\
         & 4.5          & $2454771.7598\pm0.0007$ &   $0.4976\pm0.0004$  \\
         & 8.0          & $2454771.7633\pm0.0033$ &   $0.4992\pm0.0019$  \\
\tableline
\end{tabular}
\end{center}
Note: Orbital phase for CoRoT-1 used $T_0 = 2454524.62324$ and $P=1.5089686$ days \citep{gillon09a}. 
For CoRoT-2 we used $T_0=2454237.53562$ \citep{alonso08} and $P=1.7429935$ days \citep{gillon10}.
\end{table}



\clearpage





\begin{thebibliography}{}

\bibitem[Agol et al.(2010)] {agol} Agol,~E., Cowan,~N.~B., Knutson,~H.~A., Deming,~D., 
     Steffen,~J.~H., Henry,~G.~W., \& Charbonneau,~D.  2010, ApJ, 721, 1861.

\bibitem[Alonso et al.(2008)] {alonso08} Alonso,~R., \& 42 co-authors, 2008, A\&A, 482, L21.

\bibitem[Alonso et al.(2009a)] {alonso09a} Alonso,~R., Guillot,~T., Mazeh,~T., Aigrain,~S., Barge,~P.,
    Hatzes,~A., \& Pont,~F.  2009a, A\&A, 501, L23.

\bibitem[Alonso et al.(2009b)] {alonso09b} Alonso,~R., \& 35 co-authors, 2009b, A\&A, 506, 353.

\bibitem[Alonso et al.(2010)] {alonso10} Alonso,~R., Deeg,~H.~J., Kabath,~P., \& Rabus,~M., 2010,
   AJ, 139, 1481.

\bibitem[Barge et al.(2008)] {barge} Barge,~P., \& 37 co-authors, 2008, A\&A, 482, L17.

\bibitem[Baraffe et al.(2004)] {baraffe} Baraffe,~I., Selsis,~F., Chabrier,~G., Barman,~T.~S., Allard,~F.,
    Hauschildt,~P.~H., \& Lammer,~H., 2004, A\&A, 419, L13.

\bibitem[Barman(2008)] {barman} Barman,~T., 2008, ApJ, 676, L61.

\bibitem[Bouchy et al.(2008)] {bouchy} Bouchy,~F., \& 36 co-authors, 2008, A\&A, 482, L25.

\bibitem[Burrows et al.(2007)] {burrows07} Burrows,~A., Hubeny,~I., Budaj,~J., Knutson,~H.~A., 
  \& Charbonneau,~D. 2007, ApJ, 668, L171.

\bibitem[Burrows et al.(2008)] {burrows08} Burrows,~A., Budaj,~J., \& Hubeny,~I., 2008, ApJ, 678, 1436.

\bibitem[Campo et al.(2010)] {campo} Campo,~C.~J., \& 18 co-authors, 2010, ApJ, in press, 
   astro-ph/1003.2763.

\bibitem[Charbonneau et al.(2005)]{charb05} Charbonneau,~D., Allen,~L.~E., Megeath,~S.~T.,
  Torres,~G., Alonso,~R., Brown,~T.~M., Gilliland,~R.~L., Latham,~D.~W., Mandushev,~G.,
  O'Donovan,~F., \& Sozetti,~A. 2005, ApJ 626, 523.

\bibitem[Charbonneau et al.(2008)] {charb08} Charbonneau,~D., Knutson,~H.~A., Barman,~T., Allen,~L.~E.,
  Mayor,~M., Megeath,~S.~T., Queloz,~D., \& Udry,~S., 2008, ApJ, 686, 1341.

\bibitem[Christiansen et al.(2010)] {christiansen} Christiansen,~J., Ballard,~S., Charbonneau,~D.,
  Madhusudhan,~N., Seager,~S., Holman,~M.~J., Wellnitz,~D.~D., Deming,~D., A'Hearn,~M.~F., \&
  the EPOXI Team, 2010, ApJ, 710, 97.

\bibitem[Cowan \& Agol(2010)] {cowan} Cowan,~N., \& Agol,~E., 2010, submitted to ApJ, astro-ph/1001.0012.

\bibitem[Deming et al.(2005)]{deming05} Deming,~D., Seager,~S., Richardson,~L.~J., 
  \& Harrington,~J. 2005, Nature 434, 740.

\bibitem[Deming et al.(2006)]{deming06} Deming,~D., Harrington,~J., Seager,~S., \& Richardson,~L.~J., 
  2006, ApJ, 644, 560.

\bibitem[Deming et al.(2007)] {deming07} Deming,~D., Agol,~E., Charbonneau,~D., Cowan,~N., 
  Knutson,~H., \& Marengo,~M.  2007, in {\it The Science Opportunities for the Warm Spitzer Mission},
  A.I.P. Conf. Proc., eds. L. J. Storrie-Lombardi \& N. A. Silbermann, p. 89.

\bibitem[Fazio et al.(2004)]{fazio} Fazio,~G.~G., and 64 co-authors, 2004, 
    ApJ(Suppl), 154, 10.

\bibitem[Fortney et al.(2005)] {fortney1} Fortney,~J.~J., Marley,~M.~S., Lodders,~K., Saumon,~D.,
  \& Freedman,~R.~S., 2005, ApJ, 627, L69.

\bibitem[Fortney et al.(2006)] {fortney2} Fortney,~J.~J., Saumon,~D., Marley,~M.~S., Lodders,~K.,
  \& Freedman,~R.~S., 2006, ApJ, 642, 495.

\bibitem[Fortney et al.(2008)] {fortney3} Fortney,~J.~J., Lodders,~K., Marley,~M.~S.,
  \& Freedman,~R.~S., 2008, ApJ, 678, 1419.

\bibitem[Gillon et al.(2009a)] {gillon09a} Gillon,~M., Demory,~B.-O., Triuad,~A.~H.~M.~J., 
  Barman,~T., Hebb,~L., Montalban,~J., Maxted,~P.~F.~L., Queloz,~D., Medeuil,~M., \& Magain~P.,
  2009, A\&A, 506, 359.

\bibitem[Gillon et al.(2009b)] {gillon09b}  Gillon,~M., Smalley,~B., Hebb,~L., Anderson,~D.~R.,
   Triaud,~A.~H.~M.~J., Hellier,~C., Maxted,~P.~F.~L., Queloz,~D., \& Wilson,~D.~M. 2009,
  A\&A, 496, 259. 

\bibitem[Gillon et al.(2010)] {gillon10} Gillon,~M., \& 18 co-authors, 2010, A\&A, 
   511, A3.

\bibitem[Guillot \& Havel(2010)] {guillot}  Guillot,~T. \& Havel,~M. 2010, A\&A, in press (astro-ph/1010.1078).

\bibitem[Harrington et al.(2007)] {harrington} Harrington,~J., Luszcz,~S., Seager,~S., Deming,~D.,
       \& Richardson,~L.~J., 2007, Nature, 447, 691.

\bibitem[Hauschildt et al.(1999)] {phoenix} Hauschiltd,~P.~H., Allard,~F., Ferguson,~J., Baron,~E.,
     \& Alexander,~D.~R., 1999, ApJ, 525, 871.

\bibitem[Hebb et al.(2010)] {hebb10} Hebb,~L., \& 21 co-authors, 2010, ApJ, 708, 224.

\bibitem[Hebb et al.(2009)] {hebb09} Hebb,~L., \& 34 co-authors, 2009, ApJ, 693, 1920.

\bibitem[Hebrard et al.(2010)] {hebrard} Hebrard,~G., and 26 co-authors, 2010, A\&A, 516, A95.





\bibitem[Hubbard et al.(2007)]{hubbard} Hubbard,~W.~B., Hattori,~M.~F., Burrows,~A., Hubeny,~I., \&
     Sudarsky,~D., 2007, Icarus, 187, 358. 

\bibitem[Husnoo et al.(2010)] {husnoo} Husnoo,~N., \& 11 co-authors, 2010, submitted to MNRAS,
   astro-ph/1004.1809.

\bibitem[Iro \& Deming(2010)] {Iro} Iro,~N., \& Deming,~D., 2010, ApJ, 712, 218.

\bibitem[Knutson et al.(2007)] {knutson07} Knutson,~H.~A., Charbonneau,~D., Allen,~L.~E., Fortney,~J.~J.,
    Agol,~E., Cowan,~N.~B., Showman,~A.~P., Cooper,~C.~S., \& Megeath,~S.~T., 2007, Nature, 447, 183.
  
\bibitem[Knutson et al.(2008)] {knutson08} Knutson,~H.~A., Charbonneau,~D., Allen,~L.~E., Burrows,~A., 
   \& Megeath,~S.~T. 2008, ApJ, 673, 526.

\bibitem[Knutson et al.(2009)] {knutson09} Knutson,~H.~A., Charbonneau,~D., Burrows,~A., O'Donovan,~F.~T.,
  \& Mandushev,~G. 2009, ApJ, 691, 866.

\bibitem[Knutson et al.(2010)] {knutson10} Knutson,~H.~A., Howard,~A.~W., Isaacson,~H., et al., 2010, 
   ApJ, 720, 1569.

\bibitem[Lai, Helling \& van den Heuvel(2010)]{lai} Lai,~D., Helling,~Ch., \& van~den~Heuvel,~E.~P.~J., 2010,
   ApJ, 721, 923.

\bibitem[Lanza et al.(2009)] {lanza} Lanza,~A.~F., \& 20 ao-authors, 2009, A\&A, 493, 193.

  
\bibitem[Li et al.(2010)] {li} Li,~S.-L., Miller,~N., Lin,~D.~N.~C., \& Fortney,~J.~J., 2010,
  Nature, 463, 1054.

\bibitem[Machalek et al.(2008)]{machalek} Machalek,~P., McCullough,~P.~R., Burke,~C.~J., 
   Valenti,~J.~A., Burrows,~A., \& Hora,~J.~L. 2008, ApJ, 684, 1427.



\bibitem[Morales-Calderon et al.(2006)]{morales} Morales-Calderon,~M., and 12 co-authors 2006, 
  ApJ, 653, 1454.

\bibitem[Okada et al.(2002)] {okada} Okada,~K., Aoyagi,~M., \& Iwata,~S., 2002, JQSRT, 72, 813.

\bibitem[Pont et al.(2008)] {pont} Pont,~F., Knutson,~H.~A., Gilliland,~R.~L., Moutou,~C., \& 
   Charbonneau,~D. 2008, MNRAS, 385, 109.

\bibitem[Press et al.(1992)]{press} Press,~W.~H., Teukolsky,~S.~A., Vetterling,~W.~T.,
  \& Flannery~B.~P. 1992, {\it Numerical Recipes}, Cambridge University Press.


\bibitem[Rogers et al.(2009)] {rogers} Rogers,~J.~C., Apai,~D., Lopez-Morales,~M., Sing,~D.~K., 
   \& Burrows,~A.  2009, ApJ, 707, 1707.

\bibitem[Seager, Whitney, \& Sasselov(2000)] {sws} Seager,~S., Whitney,~B.~A., 
  \& Sasselov, D.~D. 2000, ApJ, 540, 504.

\bibitem[Snellen et al.(2009)] {snellen09} Snellen,~I.~A.~G., de~Mooij,~E.J.W., \& Albrecht,~S.
   2009, Nature, 459, 543.

\bibitem[Snellen et al.(2010)] {snellen10} Snellen,~I.~A.~G., de~Mooij,~E.~J.~W., \& Burrows,~A., 
  2010, A\&A, 513, A76.

\bibitem[Spiegel et al.(2009)] {spiegel} Spiegel,~D.~S., Silverio,~K., \& Burrows,~A. 2009, ApJ,
    699, 1487.

\bibitem[Todorov et al.(2010)] {todorov} Todorov,~K., Deming,~D., Harrington,~J., Stevenson,~K.~B.,
   Bowman,~W.~C., Nymeyer,~S., Fortney,~J.~J., \& Bakos,~G.~A.  2010, ApJ, 708, 498. 




\end{thebibliography}
\end{document}